\documentclass[apj,numberedappendix]{emulateapj}
\usepackage{graphics,epsf}
\usepackage{amsmath}                
\usepackage{amsfonts}               
\usepackage{amssymb}                
\usepackage{epsfig}                 
\usepackage{graphicx}
\usepackage{float}
\usepackage{color}
\usepackage[para,online,flushleft]{threeparttable}

\begin{document}

\title{Jittering jets in cooling flow clusters}
\author{Noam Soker\altaffilmark{1}}

\altaffiltext{1}{Department of Physics, Technion, Israel; soker@physics.technion.ac.il}




\section*{}

Active galactic nucleus (AGN) jets in galactic, group, and cluster of galaxies cooling flows heat the intra-cluster medium (ICM). The heating by jets and the feeding of the AGN with cold clumps operate via a negative \emph{cold feedback mechanism} (\citealt{Soker2016} for a recent review), where jets form nonlinear perturbations that later develop into the cold clumps \citep{PizzolatoSoker2005}.

\textit{I propose that the jets not only supply energy and the perturbations for future accretion, but they influence also the angular momentum of the future accreted gas.}

Whereas there is an agreement on the cold feedback mechanism
(e.g., \citealt{ChoudhurySharma2016, Hameretal2016, Loubseretal2016, Baraietal2016, Tremblayetal2016, Donahueetal2017, Hoganetal2017, Voitetal2017, McDonaldetal2018, Voit2018}), there is no agreement on the process by which jet-inflated bubbles heat the ICM.
Suggested heating processes include excitation of shocks (e.g., \citealt{Randalletal2015}), turbulence (e.g., \citealt{Zhuravlevaetal2017}), excitation of sound waves (e.g., \citealt{Fabianetal2017}), uplifting gas (e.g., \citealt{GendronMarsolaisetal2017}), cosmic rays (e.g. \citealt{FujitaOhira2013}), and mixing (e.g., \citealt{BruggenKaiser2002, GilkisSoker2012, HillelSoker2016, YangReynolds2016b}).
Observations suggest that turbulence is too weak to heat the ICM in Perseus \citep{Hitomi2016}. This leaves the mixing of hot jet-inflated gas with the ICM the most promising heating mechanism \citep{HillelSoker2018}.

Recent observations show that jet-inflated bubbles commonly uplift cool gas from the cluster center (e.g., \citealt{Doriaetal2012, Russelletal2017, Suetal2017, GendronMarsolaisetal2017}).
Although the gravitational energy that is released by the clumps that fall back cannot be a major ICM heating source \citep{HillelSoker2018}, I show that when these clumps feed the AGN they might lead to jittering jets, i.e., the directions of the jets of different activity cycles can substantially vary.

\begin{figure}
\centering
\hskip -1.30 cm
\includegraphics[trim= 1.5cm 15cm 2cm 2.2cm,clip=true,width=0.72\textwidth]{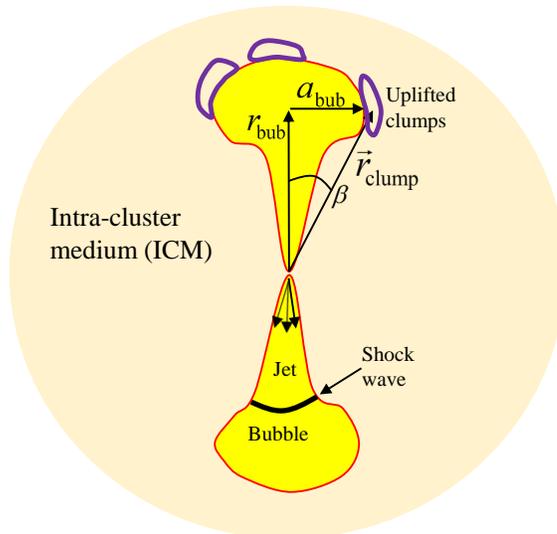}
\caption{A schematic drawing of the flow structure where jet-inflated bubbles uplift cold gas.  }
\label{fig:flow}
\end{figure}

Consider uplifted clumps (Figure \ref{fig:flow}).

At a distance of $r_{\rm bub}$ from the center the clump disconnects from the edge of the bubble. At disconnecting the clump location is $\overrightarrow{r}_{\rm clump}$ at an angle $\beta=\arctan(a_{\rm bub}/r_{\rm bub})$ from the propagation direction of the jet, and its velocity relative to the AGN is $\overrightarrow{v}_{\rm clump,0}$. Its initial specific angular momentum is $\overrightarrow{j}_{\rm clump,0}=\overrightarrow{r}_{\rm clump} \times \overrightarrow{v}_{\rm clump,0}$.
Whatever the initial velocity of the clump is, there is a minimum angle $\delta_{\rm min,c}$ between the direction of the clump's initial angular momentum and the direction of the original jet
\begin{equation}
\delta_{\rm clump} \ge \delta_{\rm min,c} = \frac{\pi}{2} -\beta =
\arctan(r_{\rm bub}/a_{\rm bub}).
\label{eq:deltamin}
\end{equation}
For Abell~1795 \citep{Russelletal2017} and Fornax  \citep{Suetal2017} I find $\delta_{\rm min,c} \approx 45^\circ$ and
$\delta_{\rm min,c} \simeq 70^\circ$, respectively.

Although most of the angular momentum of the different clumps is canceled by clump collision and friction \citep{PizzolatoSoker2010, Gasparietal2013b}, some residual angular momentum is retained by the inflowing gas.
In random clump accretion the residual angular momentum is likely to be that of the ICM. Namely, future accretion disks will tend to have the same orientation as before. In accretion from the plane of the original accretion disk, future disks will be even more likely to maintain the old orientation.

When most of the accreted cold clumps come from the direction of the original jets (Fig. \ref{fig:flow}),  equation (\ref{eq:deltamin}) shows that the residual angular momentum is likely to be at a large angle to the original one, hence varying jets' axis.

The AGN feedback heating mechanism of the ICM refers to the energy cycle. The cold feedback mechanism introduces mass as another ingredient of the feedback cycle, as the mass accretion of future AGN activity cycles is determined by the formation of perturbations by past activities. I propose that \textit{the angular momentum is also an ingredient of the feedback cycle} because the jets influence the angular momentum of future accreted gas, and hence future jets, to have varying directions. This makes the heating by mixing of the ICM more efficient, and further supports it.

\label{lastpage}

\begin{thebibliography}{}

\bibitem[Barai et al.(2016)]{Baraietal2016} Barai, P., et al.\ 2016, \mnras, 461, 1548

\bibitem[Br{\"u}ggen \& Kaiser(2002)]{BruggenKaiser2002} Br{\"u}ggen, M., \& Kaiser, C.~R.\ 2002, \nat, 418, 301

\bibitem[Choudhury \& Sharma(2016)]{ChoudhurySharma2016} Choudhury, P.~P., \& Sharma, P.\ 2016, \mnras, 457, 2554

\bibitem[Donahue et al.(2017)]{Donahueetal2017} Donahue, M., Connor, T., Voit, G.~M., \& Postman, M.\ 2017, \apj, 835, 216

\bibitem[Doria et al.(2012)]{Doriaetal2012} Doria, A., et al.\ 2012, \apj, 753, 47

\bibitem[Fabian et al.(2017)]{Fabianetal2017} Fabian, A.~C., et al.\ 2017, \mnras, 464, L1

\bibitem[Fujita \& Ohira(2013)]{FujitaOhira2013} Fujita, Y., \& Ohira, Y.\ 2013, \mnras, 428, 599

\bibitem[Gaspari et al.(2013)]{Gasparietal2013b} Gaspari, M., Ruszkowski, M., \& Oh, S.~P.\ 2013, \mnras, 432, 3401

\bibitem[Gendron-Marsolais et al.(2017)]{GendronMarsolaisetal2017} Gendron-Marsolais, M., et al.\ 2017, \apj, 848, 26

\bibitem[Gilkis \& Soker(2012)]{GilkisSoker2012} Gilkis, A., \& Soker, N.\ 2012, \mnras, 427, 1482

\bibitem[Hamer et al.(2016)]{Hameretal2016} Hamer, S.~L., et al.\ 2016, \mnras, 460, 1758

\bibitem[Hillel \& Soker(2016)]{HillelSoker2016} Hillel, S., \& Soker, N.\ 2016, \mnras, 455, 2139

\bibitem[Hillel \& Soker(2018)]{HillelSoker2018} Hillel, S., \& Soker, N.\ 2018, arXiv:1801.00408

\bibitem[Hitomi Collaboration et al.(2016)]{Hitomi2016} Hitomi Collaboration\ 2016, \nat, 535, 117

\bibitem[Hogan et al.(2017)]{Hoganetal2017} Hogan, M.~T., et al.\ 2017, \apj, 851, 66

\bibitem[Loubser et al.(2016)]{Loubseretal2016} Loubser, S.~I., et al.\ 2016, \mnras, 456, 1565

\bibitem[McDonald et al.(2018)]{McDonaldetal2018} McDonald, M., Gaspari, M., McNamara, B.~R., \& Tremblay, G.~R.\ 2018, arXiv:1803.04972

\bibitem[Pizzolato \& Soker(2005)]{PizzolatoSoker2005} Pizzolato, F., \& Soker, N.\ 2005, \apj, 632, 821

\bibitem[Pizzolato \& Soker(2010)]{PizzolatoSoker2010} Pizzolato, F., \& Soker, N.\ 2010, \mnras, 408, 961

\bibitem[Randall et al.(2015)]{Randalletal2015} Randall, S.~W., et al.\ 2015, \apj, 805, 112

\bibitem[Russell et al.(2017)]{Russelletal2017} Russell, H.~R., \ 2017, \mnras, 472, 4024

\bibitem[Soker(2016)]{Soker2016} Soker, N.\ 2016, New Astronomy Reviews, 75, 1

\bibitem[Su et al.(2017)]{Suetal2017} Su, Y., et al.\ 2017, \apj, 847, 94

\bibitem[Tremblay et al.(2016)]{Tremblayetal2016} Tremblay, G.~R., et al.\ 2016, \nat,  534, 218

\bibitem[Voit(2018)]{Voit2018} Voit, G.~M.\ 2018,  arXiv:1803.06036

\bibitem[Voit et al.(2017)]{Voitetal2017} Voit, G.~M., et al.\ 2017, \apj, 845, 80

\bibitem[Yang \& Reynolds(2016)]{YangReynolds2016b} Yang, H.-Y.~K., \& Reynolds, C.~S.\ 2016, \apj, 829, 90

\bibitem[Zhuravleva et al.(2017)]{Zhuravlevaetal2017} Zhuravleva, I., Allen, S.~W., Mantz, A.~B., \& Werner, N.\ 2017, arXiv:1707.02304

\end{thebibliography}
\end{document}